# Terahertz Pulse Shaping Using Diffractive Surfaces

Muhammed Veli[a,b,c], Deniz Mengu[a,b,c], Nezih T. Yardimci[a,b,c], Yi Luo[a,b,c], Jingxi Li[a,b,c], Yair Rivenson[a,b,c], Mona Jarrahi[a,c], Aydogan Ozcan*[abc]

[a]Department of Electrical & Computer Engineering, University of California Los Angeles (UCLA), California, USA

[b]Department of Bioengineering, University of California Los Angeles (UCLA), California, USA

[c]California NanoSystems Institute (CNSI), University of California Los Angeles (UCLA), California, USA

*E-mail: ozcan@ucla.edu

**Abstract**

Recent advances in deep learning have been providing non-intuitive solutions to various inverse problems in optics. At the intersection of machine learning and optics, diffractive networks merge wave-optics with deep learning to design task-specific elements to all-optically perform various tasks such as object classification and machine vision. Here, we present a diffractive network, which is used to shape an arbitrary broadband pulse into a desired optical waveform, forming a compact and passive pulse engineering system. We demonstrate the synthesis of various different pulses by designing diffractive layers that collectively engineer the temporal waveform of an input terahertz pulse. Our results demonstrate direct pulse shaping in terahertz spectrum, where the amplitude and phase of the input wavelengths are independently controlled through a passive diffractive device, without the need for an external pump. Furthermore, a physical transfer learning approach is presented to illustrate pulse-width tunability by replacing part of an existing network with newly trained diffractive layers, demonstrating its modularity. This learning-based diffractive pulse engineering framework can find broad applications in e.g., communications, ultra-fast imaging and spectroscopy.

## Introduction

Inspired by neural interactions in human brain[1], artificial neural networks and deep learning have been transformative in many fields, providing solutions to a variety of data processing problems, including for example image recognition[2], natural language processing[3] and medical image analysis[4]. Data-driven training of deep neural networks has set the state-of-the-art performance for various applications in e.g. optical microscopy[4–10], holography[11–16] and sensing[17–20], among others. Beyond these applications, deep learning has also been utilized to solve inverse physical design problems arising in e.g., nanophotonics and plasmonics[21–24]. These advances cover a wide range of engineering applications and have motivated the development of new optical computing architectures[25–31] that aim to benefit from the low-latency, power-efficiency and parallelization capabilities of optics in the design of machine learning hardware. For example, Diffractive Deep Neural Networks ($D^2$NN)[32] have been introduced as an optical machine learning framework that uses deep learning methods, e.g., stochastic gradient-descent and error-backpropagation, to train a set of diffractive layers for computing a given machine learning task as the light propagates through these layers. Early studies conducted on this framework showed its statistical inference capabilities, achieving >98% numerical blind testing[33,34] accuracy for the classification of the images of handwritten digits. Recently, the $D^2$NN framework has also been extended to harness broadband radiation in order to design spatially-controlled wavelength de-multiplexing systems[24]; however, this former work did not engineer the spectral phase values at different frequencies of the input radiation and therefore did not report any temporal wave control or pulse shaping.

In parallel to these recent advances at the intersection of optics and machine learning, there has been major progress in optical pulse shaping, including pulse compression for optical telecommunication[35] and pulse stretching for chirped pulse amplification[36]. Dynamic, customizable temporal waveform synthesis has been achieved using time[37–39] or frequency domain[40–42] modulation. Among different approaches, the Fourier-transform based configuration[43], which relies on conventional optical components such as lenses to establish a mapping between the pixels of an optical modulation device and the spectral components of the input broadband light, is one of the most commonly employed techniques. In various forms of its implementation, the optical modulation device placed at the Fourier plane in between two gratings can be a dynamic component e.g., a spatial light modulator[44–47], an acousto-optic modulator[48,49], a movable mirror[50] or even a metasurface[51], offering engineered dispersion and wavefront manipulation, tailored for different applications.

However, these earlier pulse shaping techniques have restricted utility at some parts of the electromagnetic spectrum, such as the terahertz band, due to the lack of advanced optical components that can provide spatio-temporal modulation and control of complex wavefronts, covering both a broad bandwidth and a high spectral resolution at these frequencies[52,53]. As a result, direct shaping of terahertz pulses by independent control of the spectral amplitude and phase of the input wavelengths has not been achieved to date; instead, the synthesis of terahertz pulses has been generally performed indirectly through the engineering of the optical-to-terahertz converters or shaping of the optical pulses that pump terahertz sources.[54–58] Previous work also demonstrated an active device using an external pump-induced inhomogeneous medium to shape input terahertz pulses.[59]

Here, we demonstrate the use of diffractive networks designed by deep learning to all-optically shape pulses by simultaneously controlling the relative phase and amplitude of each spectral component across a continuous and wide range of frequencies using only trainable diffractive layers, forming a small footprint, compact and passive pulse engineering system. This framework uses a deep learning-based physical design strategy to devise task-specific diffractive systems that can shape various temporal waveforms of interest. Following the digital training stage in a computer, we fabricated the resulting diffractive layers (Fig. 1) and experimentally demonstrated the success of our pulse shaping diffractive networks by generating pulses with various temporal widths using a broadband terahertz pulse as input. Despite using passive diffractive layers, the presented pulse shaping networks offer temporal pulse-width tunability that is experimentally demonstrated by varying the inter-layer distances within a fabricated diffractive network. We also investigated a Lego-like transfer learning approach to show the modularity of the design space provided by our framework. In addition to engineering terahertz pulses, the fundamental design approach that is presented here can be readily adapted to different parts of the electromagnetic spectrum for shaping pulses. We believe that this study extends the engineering and precise control of electromagnetic fields through deep learning-designed diffractive networks into time-domain shaping of pulses, further motivating the development of all-optical machine learning and information processing platforms that can better harness the 4D spatio-temporal information carried by light.

**Results**

**Synthesis of arbitrary temporal waveforms**

Synthesis of arbitrary temporal waveforms through small footprint and compact systems has been of great interest for various applications in e.g., tele-communications, ultra-fast imaging and spectroscopy, and it represents a challenging inverse design problem. Specifically, it requires accurate control of the complex-valued weights of the spectral components across a wide bandwidth and with high spectral resolution. We addressed this challenging inverse design problem through the training of diffractive networks as shown in Fig. 1c. The forward training model of our diffractive networks formulates the broadband light propagation using the angular spectrum representation of optical waves[24]. Based on the complex dispersion information of a diffractive material, the thickness of each diffractive feature (i.e., 'neuron') of a given diffractive layer is iteratively trained and optimized through the error-backpropagation with respect to a target cost function (see the Methods section). After the convergence of this deep learning-based training in a computer, we fabricated the resulting diffractive layers (Fig. 1c) using a 3D-printer to physically form our pulse shaping network as shown in Fig. 1a. This diffractive network was then experimentally tested for its desired/targeted pulse shaping capability using a terahertz time-domain spectroscopy (THz-TDS) setup[60] that provides a noise equivalent bandwidth of 0.1-5 THz (Figs. 1b,d).

Each one of our pulse shaping diffractive networks consists of 4 trained layers that process the input terahertz pulse to synthesize a desired temporal waveform over an output aperture of 0.2 cm × 0.2 cm. Based on this system layout and a given input pulse profile to be shaped (Fig. 2b), we trained and fabricated diffractive networks that generate square pulses with different temporal widths. For example, Figure 2a demonstrates the diffractive layers of a pulse shaping network that was trained to generate a 15.5 ps square pulse by processing the spectrum carried by the input terahertz pulse. Figure 2c demonstrates the time-domain amplitude of the output waveform numerically computed (blue) based on the trained diffractive layers and the corresponding experimentally measured temporal waveform (orange), along with the associated spectral amplitude and phase distributions. The carrier frequency of the desired temporal waveform at the output was a non-learnable, predetermined parameter set to be 0.35 THz to avoid water absorption bands in the terahertz regime (depicted by the red arrows in Fig. 2b). The numerically predicted output waveform (blue) in Fig. 2c indicates that a 4-layer diffractive network can synthesize a square temporal waveform with a pulse width of 15.69 ps without using any conventional optical components, in a compact architecture that spans approximately 250-times the carrier wavelength in the axial direction. The pulse width of the temporal waveform created by the 3D printed diffractive layers at the output aperture is measured as 15.52 ps, closely matching the numerically predicted result (15.69 ps). Similarly, a comparison of the output spectral amplitude profiles for the numerical and experimental results shows a good agreement in terms of the peak locations of the main and side lobes as well as the relative amplitude carried by each spectral component. On the other hand, an examination of the unwrapped phase profiles (experimental vs. numerical) reveals that the 3D-fabricated, physical diffractive network could not exactly create the sharp phase transitions at the expected spectral locations, but rather generated smoothened transitions. This smoothening contributes to some of the differences observed between the experimentally measured and the numerically calculated time-domain waveforms (Fig. 2c). The power efficiency of this diffractive network was experimentally measured as ~0.51% at the carrier frequency ($f_0$ = 355 GHz), quantified at the output aperture, normalized with respect to the input; here we should emphasize that >70% of the input optical power at the carrier frequency is in fact lost due to absorption within the 3D printed diffractive layers. Therefore, to create our diffractive layers, the selection of a different fabrication material with a much lower loss (e.g., polymers such as poly-methylpentene: TPX)[61–63] can significantly boost the overall efficiency of these diffractive pulse shaping networks. Other strategies to improve the power efficiency include increasing the output aperture size and introducing additional power-related penalty terms during the training phase of the diffractive network (see the Discussion section).

Supplementary Figure 1 further illustrates another diffractive network that was designed to create a narrower square pulse at its output aperture. At the end of its deep learning-based training, the numerical forward model converged to the thickness profiles shown in Supplementary Fig. 1a in order to synthesize a 10.96 ps square pulse (blue) illustrated in Supplementary Fig. 1c. When the diffractive layers depicted in Supplementary Fig. 1a were 3D printed and experimentally tested using the setup shown in Fig. 1d, the output pulse waveform was measured to have a temporal width of 11.85 ps (orange curve in Supplementary Fig. 1c), providing a good match to our numerical results, similar to the conclusions reported in Fig. 2.

Beyond fabrication artefacts and misalignments observed in the 3D-printed diffractive networks, the variation of the input terahertz pulse from experiment to experiment is one of the significant contributors for any mismatch between the numerical and experimental output waveforms. The deep learning-based design of the diffractive networks shown in Fig. 2 and Supplementary Fig. 1 relies on a known input terahertz pulse profile that is experimentally measured over the input aperture. To be able to take into account uncontrolled variations of the input pulse profile from run to run, we used 5 different experimentally measured input pulse profiles (dashed curves in Supplementary Figs. 2a-b) during the training phase of each diffractive network. In the experimental testing phase, however, the terahertz input pulse (light blue curve in Supplementary Figs. 2a-b) slightly deviated from these input pulse profiles used in the training, causing some distortions in the experimental results shown in Fig. 2 and Supplementary Fig. 1, compared to their numerically computed counterparts for the same diffractive network models (also see Supplementary Fig. 3 and Supplementary Fig. 4).

To shed more light onto this, next we normalized the experimentally measured spectral amplitude profiles depicted in Fig. 2c and Supplementary Fig. 1c, based on the ratio between the average spectral amplitudes carried by the input pulses used in the training phase and the input pulse measured at the experimental testing phase. This simple spectral normalization procedure nullifies the effect of input terahertz source variations from experiment to experiment and provides us an opportunity to better evaluate the accuracy of the complex-valued spectral filtering operation performed by the 3D-fabricated diffractive network. Supplementary Figs. 2c and 2d demonstrate the experimental spectral amplitudes and the corresponding temporal waveforms at the network output before and after this spectral normalization step for the diffractive networks shown in Fig. 2a and Supplementary Fig. 1a, respectively. Following the spectral normalization, the width of the square pulse created by the diffractive network in Supplementary Fig. 1a, for example, decreased from 11.85 ps to 10.49 ps, providing a better match to the 11.07 ps that is predicted by our numerical forward model (Supplementary Fig. 2d). A similar improvement using spectral normalization was also observed for the diffractive network shown in Fig. 2a, almost perfectly matching its numerical counterpart in terms of the square pulse width, achieving 15.71 ps after the normalization step (Supplementary Fig. 2c).

These results highlight that experiment-to-experiment variability of our input terahertz pulse profile causes it to deviate from the input pulse profiles used in the training phase of our diffractive network, creating some uncontrolled errors in the output pulse profile, which can be improved significantly after the spectral normalization step, as discussed above. To further explore the pulse shaping capabilities of diffractive networks, next we trained a set of generic diffractive networks that used/assumed a flat input spectrum during their training in order to achieve a desired output waveform; stated differently, a generic diffractive network is trained using an input pulse where all the wavelengths have the same spectral amplitude and phase. To accurately demonstrate the pulse shaping behavior of these generic diffractive designs that were trained with flat spectra, we used spectral normalization based on the input pulse profile, experimentally measured at each run. For example, Supplementary Fig. 5a and 6 show the diffractive layers of a generic pulse shaping network model that was trained to create a 15.5 ps square pulse. Supplementary Fig. 5c reports the time-domain amplitude of the output waveform numerically computed (blue) based on these trained diffractive layers and the experimentally measured temporal waveform (orange) along with the corresponding spectral amplitude and phase distributions. The synthesized pulse shape by the 3D-printed diffractive network closely matches the numerically computed waveform using our forward model, despite the water absorption bands that appear in our experimental results, illustrated by the red arrows in Supplementary Fig. 5b. The power efficiency at the carrier frequency ($f_0 = 400$ GHz) of this diffractive network was experimentally measured as ~0.97%. Figure 3 further demonstrates three additional generic pulse shaping diffractive network models that were trained with a flat input spectrum and experimentally tested using our terahertz setup to achieve different square pulses, with pulse widths of 11.25 ps, 13.45 ps and 16.69 ps, respectively, demonstrating a very good match to their numerical counterparts. The numerically computed peak frequencies for these three different diffractive networks were 399.4 GHz, 396.1 GHz and 399.4 GHz, which were measured experimentally as 399.1 GHz, 402.2 GHz and 401.8 GHz, respectively. As we move towards higher optical frequencies beyond 0.6 THz, the experimental spectral amplitude distributions start to deviate from their numerically predicted counterparts. Considering that the maximum material thickness in our model is

~1mm, at higher optical frequencies corresponding to wavelengths below ~0.5 mm, the light may travel more than 2 wavelengths inside a diffractive feature (depending on the final trained model) which will then violate the thin modulation layer assumption in our forward model contributing to some of the experimental errors observed in Fig. 3. In addition, the size each diffractive feature corresponding to a unique complex-valued modulation per neuron (see Methods) was chosen to be 0.5 mm due to the limited lateral resolution of our 3D printer. Therefore, for higher frequencies, the light fields are modulated at each diffractive layer with 2D functions sampled at lower spatial rates, which, in return, partially limits the design capabilities of our diffractive networks at those smaller wavelengths of the pulse bandwidth. Furthermore, the uneven surface profile in 3D printing combined with thickness variations induced by fabrication imperfections contribute to some additional sources of experimental errors observed in our results.

To further demonstrate the design capabilities of our diffractive pulse shaping framework, in addition to the square pulses with various temporal widths reported earlier, we also trained three new diffractive network models that were designed to output (1) a chirped-Gaussian pulse (Supplementary Fig. 7), (2) a sequence of positive and negative chirped Gaussian pulses, one following another (Supplementary Fig. 8) and (3) a sequence of two chirp-free Gaussian pulses (Supplementary Fig. 9). These results report a very good match, both in time and spectral domains, between the target, ground-truth pulse profiles and the corresponding output pulses synthesized by the trained diffractive networks, clearly demonstrating the versatile nature of the presented framework to synthesize arbitrary pulses, engineered through the deep learning-based design of diffractive surfaces.

**Pulse width tunability**

Next, we demonstrated the temporal width tunability of pulse shaping diffractive networks despite the passive nature of their layers. By changing the axial distance between successive diffractive layers by $\Delta Z$, the temporal width and the peak frequency of the output waveform can be tuned without any further training or a change to the 3D printed diffractive layers. We demonstrated this pulse-width tunability using the 3D printed diffractive network depicted in Supplementary Fig. 5, but a similar tunability also applies to the network models shown in Fig. 3. Since our diffractive networks used 30 mm layer-to-layer distance in their design, we considered the $\Delta Z$ range to be between -10 mm to 20 mm; for instance, when $\Delta Z$ is taken as -10 mm, the axial distance between all the successive layers of the diffractive network is set to be 20 mm. Within this axial tuning range, Figs. 4a-h demonstrate the effect of changing this layer-to-layer distance of an already designed/trained diffractive network on the output waveform and its complex-valued spectrum. The results reveal that as the diffractive layers get closer to each other axially, i.e., a negative $\Delta Z$, the pulse-width of the output waveform increases and the peak frequency decreases. For instance, when the axial distance between each diffractive layer of the design shown in Supplementary Fig. 5 is decreased by 5 mm (ΔZ = -5 mm) as shown in Fig. 4d, the peak of the spectral amplitude distribution shifts from 399.4 GHz to 349.1 GHz according to our numerical forward model. The pulse-width of the resulting square pulse at the output aperture was numerically found to be 17.59 ps suggesting a longer pulse compared to 15.56 ps synthesized by the original design, ΔZ = 0 mm (Fig. 4d). The experimentally measured pulse width with the same amount of axial change in the layer-to-layer distance of the diffractive network revealed a 17.56 ps pulse after the spectral normalization step, confirming the tunability of our pulse shaping diffractive network and also providing a very good match to our numerical results (Fig. 4).

When the layer-to-layer distance is increased, i.e., a positive ΔZ, the output square pulse gets narrower in time domain with an accompanying shift in the peak frequency toward higher values. Figure 4e demonstrates an example of this case with ΔZ = 5 mm, i.e. the distance between each diffractive layer is increased to 35 mm. In this case, the experimentally measured and numerically computed square pulses at the output plane have peak frequencies of 451.4 GHz and 453.1 GHz, with the corresponding pulse-widths of 14.3 ps and 13.97 ps, respectively, once again confirming the tunability of our pulse shaping diffractive networks and demonstrating a very good agreement between the numerical forward model and our experiments. As we further increase ΔZ beyond 10 mm (depicted in Fig. 4f), the time domain pulse continues to get narrower.

**Modularity of diffractive pulse shaping network**

To further explore methods to alter a given fabricated diffractive network and its output function, next we employed a Lego-like physical transfer learning approach to demonstrate pulse-width tunability by updating only part of a pre-trained network with newly trained and fabricated diffractive layers, showing the modularity of a diffractive pulse shaping network. For this aim, we took the pre-trained network that experimentally synthesized a 15.57 ps square waveform, noted as the original design in Fig. 5a, and further trained only the last diffractive layer to synthesize a new desired output waveform, i.e., a 12.03 ps square pulse, by keeping the first three layers as they are (already fabricated). We experimentally validated this transfer learning approach as shown in Fig. 5b by removing the existing last diffractive layer and inserting a newly trained layer, fabricated using the same 3D printer. Numerical and experimental results revealed very good match to each other for the normalized output spectral amplitude over a wide frequency range as well as for the normalized output field waveform, generating pulse-widths of 12.21 ps and 13.25 ps, respectively. Next we took an alternative approach: this time, the last two diffractive layers were replaced with new diffractive layers trained to generate 12.03 ps square pulses. As illustrated in Fig. 5c, with the addition of these two new diffractive layers to the already existing first two layers, the resulting new diffractive network successfully demonstrated the synthesis of 12.14 ps and 12.39 ps waveforms at the output aperture for the numerical and experimental waveforms, respectively. The peak frequency of the new network model was calculated to be at 399.4 GHz and it was experimentally measured to be at 399.8 GHz, showing once again a very good match between our numerical forward model and experimental results. Overall, the insertion of two newly trained layers, when compared to a single newly trained layer added on top of the existing layers of a 3D-fabricated network, provided us improved performance for achieving the new pulse form that is desired.

**Discussion**

Our results reported in earlier sub-sections demonstrate direct pulse shaping in terahertz part of the spectrum, where a complex-valued spectral modulation function that is trained using deep learning directly acts on terahertz frequencies through a passive diffractive device, without the need for an external pump. The presented learning-based approach can shape any input terahertz pulse through diffraction and is fundamentally different from previous approaches that indirectly synthesize a desired terahertz pulse through optical-to-terahertz converters or shaping of the optical pump that interacts with terahertz sources. This capability of direct pulse shaping in terahertz band enables new opportunities that could not be explored with indirect pulse shaping approaches. For example, precise engineering and synthesis of terahertz pulses with the state-of-the-art methods is either not possible or very hard and costly to achieve, including e.g., pulsed terahertz generation through quantum cascade lasers[64–66], solid-state circuits[67,68] and particle accelerators[69]. Furthermore, the presented deep learning-based framework is quite flexible and versatile that can be used to engineer terahertz pulses regardless of their polarization state, beam shape, beam quality or aberrations of the specific terahertz generation mechanism.

The intrinsic pulse-width tunability of a given diffractive network that is achieved by changing the axial layer-to-layer distance is an interesting feature that we demonstrated numerically and experimentally: Figure 4a shows various pulse-widths obtained at seven different layer-to-layer distances using an existing network design. As the layer-to-layer distance of a diffractive network design increases, the temporal pulse-width at the output aperture gets smaller, without any further training or fabrication of new diffractive layers. This opens up the opportunity to synthesize new waveforms within a certain time window around the originally designed output pulse. In addition to that, an axial distance change between the existing layers of a diffractive network also shifts the center frequency of the output pulse as shown Figure 4b. As the diffractive layers get closer to each other, we observed a red-shift in the center frequency. Another related aspect of this pulse shaping diffractive framework is its modularity to tune the output pulses using a Lego-like physical transfer learning approach. By training a new layer (or layers) to replace part of an existing, pre-trained diffractive network model, on demand synthesis of new pulses can be achieved, as demonstrated in Figures 5b-c. These results highlight some of the unique features of diffractive pulse shaping networks and how they can adapt to potential changes in the desired output pulse patterns.

The presented pulse shaping framework has a compact design, with an axial length of approximately $250 \times \lambda_0$, where $\lambda_0$ denotes the peak wavelength. Moreover, it does not utilize any conventional optical components such as spatial light modulators, which makes it ideal for pulse shaping in terahertz part of the spectrum, where high-resolution spatio-temporal modulation and control of complex wavefronts over a broad bandwidth represent a significant challenge. In addition to being compact and much simpler compared to previous demonstrations of pulse shaping in terahertz spectrum, our results present the implementation of direct pulse shaping in terahertz band, where the learned complex-valued spectral modulation function of the diffractive network directly acts on terahertz frequencies for pulse engineering. This capability enables new opportunities: when merged with appropriate fabrication methods and materials, the presented pulse shaping approach can be used to directly engineer terahertz pulses generated through quantum cascade lasers, solid-state circuits and particle accelerators. Another major advantage of this deep learning-based approach is that it is versatile and can be easily adapted to engineer terahertz pulses irrespective of their polarization state, beam quality as well as spectral/spatial aberrations.

The experimentally measured power efficiency values reported in our manuscript are ~1%. However, there are various design strategies that can increase power-efficiency in diffractive pulse shaping networks as detailed in table in Fig. 6 (also see the Methods section). The diffractive networks reported in Fig. 6 were trained to synthesize 15.5 ps square pulses at their output plane. As one can observe in Fig. 6, the power efficiency values of the resulting diffractive models can be increased by more than an order of magnitude by adjusting the training loss function, increasing the output aperture size and using low absorption materials. For example, as reported in the second column of table in Fig. 6, when the material absorption is ignored during the testing of a diffractive network model, a 2-fold wider output aperture (i.e., 4 mm) provides a significant improvement in the power efficiency of the pulse shaping networks, reaching 60.37% and 61% for two different network models. On the other hand, if the absorption of our 3D-printing material is taken into account as part of the optical forward model, one can reach an efficiency value of 17.84% by accordingly optimizing the training loss function and using a 4 mm output aperture (see Fig. 6).

By comparing the top and bottom efficiency values for a given training loss function and design strategy reported in Fig. 6, we clearly see that the 3D-printing material used in this work decreases the pulse shaping network efficiency 2-5 times, in different designs, compared to an ideal, non-absorbing optical material. As an alternative fabrication material for diffractive pulse shaping networks, one can consider low-absorption polymers[61–63] used in commercially available components designed for THz wavelengths, such as TPX, which exhibits a two orders-of-magnitude smaller absorption coefficient compared to the 3D printing material used in our work. There have been various fabrication processes developed for such low absorption polymers[70,71], which can be used to precisely control the thickness of these low-loss polymers with a relatively high-resolution to manufacture pulse shaping diffractive networks with much lower material absorption. To even further improve the output efficiency of pulse shaping diffractive networks, anti-reflective (AR) coatings over diffractive surfaces can also be utilized to reduce back-reflections, similar to the AR-coated commercial lenses and other optical components.

In conclusion, we presented a modular pulse shaping network that synthesizes various pulse waveforms using deep learning. Precise shaping of the spectral amplitude and phase profile of an arbitrary input pulse over a wide frequency range can be achieved using this platform, which will be transformative for various applications including e.g., communications, pulse compression, ultra-fast imaging and spectroscopy. In addition to direct engineering of terahertz pulses, the presented diffractive pulse shaping network can be utilized in different parts of the electromagnetic spectrum by using appropriate fabrication technologies and materials.

**Methods**

**Terahertz setup**

Figure 1 shows the schematic diagram of the terahertz time-domain spectroscopy (THz-TDS) setup that was used to measure the input and output pulse profiles reported in this work. A Ti:sapphire laser (Coherent Mira HP) is used to generate femtosecond optical pulses. The optical beam generated by the laser is split into two parts. One part of the beam is used to pump a high-power plasmonic photoconductive terahertz source to generate terahertz pulses[72], which are collimated with off-axis parabolic mirrors and guided to a high-sensitivity plasmonic photoconductive terahertz detector[60]. The other part of the beam passes through an optical delay line (Newport IMS300LM) and is focused onto the terahertz detector. As a result, an ultrafast signal which is directly proportional to the incident terahertz field is generated within the terahertz detector. The signal is sampled with a 12.5 fs time-resolution over a 400 ps time-window by changing the time delay between the terahertz and optical probe pulses incident on the detector, amplified with a transimpedance pre-amplifier (Femto DHPCA-100), and acquired with a lock-in amplifier (Zurich Instruments MFLI). For each measurement, 10 time-domain traces are collected and averaged. The described THz-TDS setup provides a 90 dB signal-to-noise ratio over a 5 THz noise-equivalent-power bandwidth.

Each one of the pulse shaping diffractive networks consists of 4 trained layers that are separated by 3 cm as illustrated in Fig. 1. The diffractive layers, input and output apertures, were fabricated using a 3D Printer (Objet30 Pro, Stratasys Ltd.). The fabrication/preparation of each diffractive layer takes approximately 1.5-2 hours. A square input aperture (0.8 cm) and an output aperture (0.2 cm) are placed 3 cm from the first diffractive layer and 10 cm from the last diffractive layer, respectively (Fig. 1c). The printed apertures were aluminum coated to prevent any light wave passing through the regions outside of the aperture. After the design and printing of the diffractive layers, they were placed at their corresponding locations inside a 3D printed holder that ensures robust alignment between the layers. During the pulse shaping experiments, the diffractive network was directly placed between the terahertz source and detector, coaxial with the terahertz input pulse emanating from the source (Figs. 1b,d). After the alignment of the diffractive network, the output pulse was measured and it was followed by the measurement of the reference input pulse which was acquired by placing the same terahertz detector at the input aperture, without any diffractive layers between the source and detector. For generic diffractive networks that were trained with flat input spectra, the measured output pulse spectrum is normalized with respect to the measured reference input pulse and its spectral amplitude is smoothened around water absorption lines shown in Figs. 3-5 and Supplementary Fig. 5. The measured pulse width at the network output is defined as the width of the time interval that the envelope of the pulse amplitude is at least 20% of its maximum (see e.g., Figs. 2-5).

**Forward Model**

Our forward model considers the layers of a diffractive network as thin modulation elements that are connected to the next layer through free space propagation. The modulation of neurons at each layer can be modeled as:

$$M^n(x_i, y_i, z_i, \lambda) = A^n(x_i, y_i, z_i, \lambda) \exp\left(j\varphi^n(x_i, y_i, z_i, \lambda)\right) \quad (1),$$

where $M$ represents the complex transmission/reflection coefficient. The field amplitude, phase, wavelength, and diffractive layer number are denoted by $A$, $\phi$, $\lambda$ and $n$ respectively. Free space propagation between each layer is calculated based on the Rayleigh-Sommerfeld formulation of diffraction that models a diffractive feature as source of a secondary wave:

$$W_i^n(x, y, z, \lambda) = \frac{z-z_i}{r^2}\left(\frac{1}{2\pi r} + \frac{1}{j\lambda}\right) \exp\left(\frac{j 2\pi r}{\lambda}\right) \quad (2),$$

where $r = \sqrt{(x-x_i)^2 + (y-y_i)^2 + (z-z_i)^2}$, $j = \sqrt{-1}$ and $W_i^n(x, y, z, \lambda)$ is the secondary wave generated by the i[th] neuron on n[th] layer at location $(x_i, y_i, z_i)$, respectively. Then, we can write the optical field at layer n, at point $(x_i, y_i, z_i)$ as:

$$U^n(x_i, y_i, z_i, \lambda) = M^n(x_i, y_i, z_i, \lambda) \sum_k U^{n-1}(x_k, y_k, z_k, \lambda) W_k^{n-1}(x_i, y_i, z_i, \lambda), n \geq 1 \quad (3).$$

**Network Training**

During the training of a pulse shaping diffractive network, one of the 5 pulses measured at the input plane (Supplementary Figs. 2a-b) were randomly selected as the input pulse at each iteration of the training model for the diffractive networks reported in Fig. 2 and Supplementary Fig. 1; for the generic diffractive network models reported in Figs. 3-5, Supplementary Fig. 5 and Supplementary Figs. 7-9, however, the input is modeled as a spectrally flat Gaussian beam with varying FWHM values over a wide frequency range (Supplementary Fig. 10) and with a uniform phase profile. The wave propagation is performed for N = 300 discrete frequencies that were uniformly sampled between 3 GHz and 1 THz.

In our wave propagation through the diffractive layers, a 0.5 mm pixel (i.e., diffractive feature) size is assumed based on the lateral resolution of our 3D printer. While a pixel size of 0.5 mm can create all the propagating modes of free-space for frequencies below ~300 GHz, they can only excite plane waves over a subset of the k-vectors supported by the free-space for the spectral components between 300 GHz and 1 THz.[73] Therefore, diffractive pulse shaping networks would in general benefit from higher resolution fabrication techniques with better lateral resolution to more accurately control and engineer the complex-valued spectral weights of a given desired pulse.

To calculate the Rayleigh-Sommerfeld integral more accurately, each pixel is oversampled twice so that all 4 elements have the same thickness values in that 2×2 grid. The thickness of each pixel, $h$, is composed of a base height ($h_{base}$) of 0.1 mm, which provides adequate mechanical stiffness to the fabricated diffractive layer and a trainable modulation height ($h_{tr}$) that is between 0 and 1 mm, i.e.,

$$h = h_{base} + h_{tr} \quad (4)$$

To confine the modulation height between 0 and 1 mm, we defined $h_{tr}$ over an auxiliary training-related variable, $h_a$, using:

$$h_{tr} = 0.5mm \times \{1 + \sin(h_a)\} \quad (5),$$

In its general form, the amplitude and phase modulation of each neuron of a given diffractive layer is a function of the layer thickness, incident wavelength, material extinction coefficient $\kappa(\lambda)$ and refractive index $n(\lambda)$, i.e.,

$$A^n(x, y, z, \lambda) = \exp\left(-\frac{2\pi\kappa(\lambda)h}{\lambda}\right) \quad (6)$$

$$\varphi^n(x, y, z, \lambda) = \frac{2\pi h(n(\lambda) - n_{air})}{\lambda} \quad (7)$$

The material refractive index $n(\lambda)$ and the extinction coefficient $\kappa(\lambda)$ are defined as the real and imaginary parts of the complex refractive index, $\tilde{n}(\lambda) = n(\lambda) + j\kappa(\lambda)$, determined by the dispersion of our 3D fabrication material[24]. Since we have relatively small variations in the extinction coefficient over the frequency band that we utilized in this work, we ignored the material absorption during the training and numerical simulations of diffractive layers.

After the wave propagation through diffractive layers, light goes through the output square aperture of 2mm width, which is placed right in front of the hemisphere silicon lens which is 1.2 cm in diameter. Since the effective aperture of this Si lens was significantly restricted by the output aperture, it was modeled as a uniform slab with a refractive index of 3.4 and 0.5 cm thickness. After the propagation through the Si slab, the coherent integration of the optical waves incident on the active area of the detector was computed to obtain the spectral field amplitude and phase for each frequency. The power efficiency was defined as $\eta_{f_0} = \frac{I_{sensor,f_0}}{I_{input,f_0}}$ for the peak/center frequency ($f_0$) of given diffractive network design, where $I_{input,f_0}$ and $I_{sensor,f_0}$ denote the power within the input and output apertures, respectively.

Our loss function ($L$) used during the training phase has three components: temporal loss term ($L_t$) which penalizes the mismatch between the target and the output time waveforms, the power loss term ($L_p$), and the power surrounding the detector region ($L_s$), i.e.,

$$L = \alpha L_t + \beta L_p + L_s \quad (8)$$

To calculate the temporal loss, $L_t$, first the output temporal waveform is reconstructed from the spectral field amplitude and phase on the detector area, and it is normalized. Then, the difference between the target temporal waveform and the reconstructed output waveform is integrated over time:

$$L_t = \sum_t \left(f_{\text{target}} - f_{\text{output}}\right)^2 \quad (9),$$

where $f_{\text{target}}$ and $f_{\text{output}}$ denote the ground-truth, time-domain waveform and the synthesized waveform by the diffractive network model at a training iteration. For a given diffractive network model, $f_{output}$ is computed by propagating the input waves of all the spectral components from the input aperture to the output aperture. Next, the complex-valued wave fields of these different wavelength components are integrated over the sensitive area of the detector to obtain each complex-valued spectral coefficient at the output, which is followed by an inverse Fourier transform operation over the resulting vector. Alternatively, the error term between a target, time-domain pulse, $f_{\text{target}}$, and the synthesized waveform by the diffractive network, $f_{\text{output}}$, can directly be computed based on the complex-valued spectral coefficients without any inverse Fourier transform operation. However, in this case, since the error is defined based on the complex-valued target and output functions, two separate error functions must be computed for the real and imaginary parts of the spectral coefficients and these two losses must be combined to compute the final loss term.

The analytical form of the square pulses used in this work can be written as: $f_{\text{target}}(t) = rect(bt)\cos(2\pi f_0 t)$, where $f_0$ and $b$ represent the carrier frequency and the rectangular pulse-width, respectively. For the Gaussian pulses, however, the analytical form of the target waveform can be written as: $f_{\text{target}}(t) = \sum_{i=1}^{n} C_i \cos(2\pi f_0(t - t_{0,i})) \exp\left(-(t - t_{0,i})^2 / (2p_i)\right) \exp(jq_i(t - t_{0,i})^2)$, where $t_{0,i}$, $C_i$, $p_i$ and $q_i$ denote the time instant of the peak, magnitude, variance of the low-pass envelope and the instantaneous angular chirpiness, respectively. The number of desired pulses inside a targeted time-window is determined by $n$. For the examples shown in Supplementary Figs. 7-9, the target time domain waveforms were created by setting these parameters to [$n$ = 1, $t_0$ = 0, $C_1$ = 1, $p_1$ = 2.2×10$^{-22}$, $q_1$ = 5.76×10$^{21}$]; [$n$ = 2, $t_{0,1}$ = 0, $t_{0,2}$ = 27 ps, $C_1$ = 1, $C_2$ = 0.5, $p_1$ = $p_2$ =1.38×10$^{-23}$, $q_1$ = 6.25×10$^{22}$ $q_2$ = -6.25×10$^{22}$]; and [$n$ = 2, $t_{0,1}$ = 0, $t_{0,2}$ = 19 ps, $C_1$ = 1, $C_2$ = 1, $p_1$ = $p_2$ =4.58×10$^{-24}$, $q_1$ = $q_2$ = 0], respectively.

For the diffractive network designs shown in the last row of the table in Fig. 6, we used a power loss term, $L_p$, defined as:

$$L_p = \begin{cases} -\log\left(\dfrac{\eta}{\eta_{\text{th}}}\right), & \text{if } \eta < \eta_{\text{th}} \\ 0, & \text{if } \eta \geq \eta_{\text{th}} \end{cases} \quad (10),$$

where $= \dfrac{\sum_\omega I_{\text{sensor}}}{\sum_\omega I_{\text{input}}}$. $I_{\text{input}}$ and $I_{\text{sensor}}$ denote the power within the input and output apertures for a given wavelength, respectively. For the diffractive network designs shown in the last row of the table in Fig. 6, corresponding to 2 mm and 4 mm output apertures, $\eta_{\text{th}}$ was selected as 0.07 and 0.08, respectively. For the all remaining designs reported in the manuscript, the power loss term is defined as:

$$L_p = \frac{\sum_\omega (I_{\text{target}} - I_{\text{sensor}})^2}{\sum_\omega I_{\text{target}}^2} \tag{11}$$

where $I_{\text{target}}$ is the total power of the target waveform at a given wavelength within the input aperture, normalized with respect to the power of the input at the center frequency, $f_0$.

The last component of our loss function which represents the power surrounding the detector aperture is defined as:

$$L_s = \frac{\sum_\omega I_{\text{surround}}}{\sum_\omega I_{\text{output plane}}} \tag{12}$$

where $I_{\text{surround}}$ is the total power at a given wavelength within the 5 mm × 5 mm square region that is centered around the output aperture (excluding the output aperture, i.e., it only measures the signal surrounding the output aperture) and $I_{\text{output plane}}$ is the total power at a given wavelength within the output plane.

The diffractive networks that synthesized 10.58 ps, 10.96 ps, 13.26 ps, 15.56 ps, 15.69 ps and 17.94 ps square terahertz pulses were trained with $\frac{\alpha}{\beta}$ ratios of 6500, 500, 4500, 1500, 750000 and 2500, respectively. For the Lego-like transfer learning approach, an $\frac{\alpha}{\beta}$ ratio of 8500 was used. For Supplementary Figs. 7 and 9, we used an $\frac{\alpha}{\beta}$ ratio of 1500, and for Supplementary Fig. 8, we used $\frac{\alpha}{\beta} = 15000$.

Figure 6 reports a series of diffractive optical network designs that are trained to create a square pulse of 15.5 ps at their output apertures, achieving different levels of power efficiencies. Among these pulse shaping diffractive network models, the $\frac{\alpha}{\beta}$ ratio was adjusted depending on the size of the output aperture. Specifically, the diffractive networks targeting a 2mm aperture at the output plane were trained with $\frac{\alpha}{\beta} = 1500$, and this ratio was reduced to 136 for the diffractive pulse shaping systems with 4mm wide output apertures. Finally, an $\frac{\alpha}{\beta}$ ratio of 150 was used for the diffractive optical networks that were trained with the power efficiency loss term described in Eq. 10.

In our training, Adam optimizer is used as a standard error backpropagation method with a learning rate of $0.8 \times 10^{-3}$ for the pulses reported in Supplementary Figs. 7-9. For the diffractive networks synthesizing 10.96 ps and 15.69 ps square pulses, on the other hand, the learning rate was set to be $10^{-3}$. For the rest of the diffractive network models $10^{-4}$ was used as the learning rate. All the trainable parameters were initialized as zero. Our designs used Python (v3.7.3) and TensorFlow (v1.15.0) on a computer that has Nvidia Titan RTX graphical processing unit, Intel

Core i9 CPU and 128 GB of RAM with Windows 10 operating system. MATLAB 2016b is used to convert designed diffractive layers to a 3D printable (.stl) file format.

**References**

1   Cox DD, Dean T. Neural Networks and Neuroscience-Inspired Computer Vision. *Current Biology* 2014; **24**: R921–R929.

2   LeCun Y, Bengio Y, Hinton G. Deep learning. *Nature* 2015; **521**: 436–444.

3   Collobert R, Weston J. A unified architecture for natural language processing: deep neural networks with multitask learning. In: *Proceedings of the 25th international conference on Machine learning*. Association for Computing Machinery: Helsinki, Finland, 2008, pp 160–167.

4   Litjens G, Kooi T, Bejnordi BE, Setio AAA, Ciompi F, Ghafoorian M *et al.* A survey on deep learning in medical image analysis. *Medical Image Analysis* 2017; **42**: 60–88.

5   Rivenson Y, Ceylan Koydemir H, Wang H, Wei Z, Ren Z, Günaydın H *et al.* Deep Learning Enhanced Mobile-Phone Microscopy. *ACS Photonics* 2018; **5**: 2354–2364.

6   Rivenson Y, Göröcs Z, Günaydin H, Zhang Y, Wang H, Ozcan A. Deep learning microscopy. *Optica, OPTICA* 2017; **4**: 1437–1443.

7   Nehme E, Weiss LE, Michaeli T, Shechtman Y. Deep-STORM: super-resolution single-molecule microscopy by deep learning. *Optica, OPTICA* 2018; **5**: 458–464.

8   Ouyang W, Aristov A, Lelek M, Hao X, Zimmer C. Deep learning massively accelerates super-resolution localization microscopy. *Nature Biotechnology* 2018; **36**: 460–468.

9   Wang H, Rivenson Y, Jin Y, Wei Z, Gao R, Günaydın H *et al.* Deep learning enables cross-modality super-resolution in fluorescence microscopy. *Nature Methods* 2019; **16**: 103–110.

10   Wu Y, Rivenson Y, Wang H, Luo Y, Ben-David E, Bentolila LA *et al.* Three-dimensional virtual refocusing of fluorescence microscopy images using deep learning. *Nature Methods* 2019; **16**: 1323–1331.

11   Rivenson Y, Zhang Y, Günaydın H, Teng D, Ozcan A. Phase recovery and holographic image reconstruction using deep learning in neural networks. *Light: Science & Applications* 2018; **7**: 17141–17141.

12   Rivenson Y, Liu T, Wei Z, Zhang Y, de Haan K, Ozcan A. PhaseStain: the digital staining of label-free quantitative phase microscopy images using deep learning. *Light: Science & Applications* 2019; **8**: 23.

13      Wu Y, Ray A, Wei Q, Feizi A, Tong X, Chen E *et al.* Deep Learning Enables High-Throughput Analysis of Particle-Aggregation-Based Biosensors Imaged Using Holography. *ACS Photonics* 2019; **6**: 294–301.

14      Sinha A, Lee J, Li S, Barbastathis G. Lensless computational imaging through deep learning. *Optica, OPTICA* 2017; **4**: 1117–1125.

15      Wu Y, Luo Y, Chaudhari G, Rivenson Y, Calis A, de Haan K *et al.* Bright-field holography: cross-modality deep learning enables snapshot 3D imaging with bright-field contrast using a single hologram. *Light: Science & Applications* 2019; **8**: 25.

16      Wu Y, Rivenson Y, Zhang Y, Wei Z, Günaydin H, Lin X *et al.* Extended depth-of-field in holographic imaging using deep-learning-based autofocusing and phase recovery. *Optica, OPTICA* 2018; **5**: 704–710.

17      Ballard ZS, Joung H-A, Goncharov A, Liang J, Nugroho K, Carlo DD *et al.* Deep learning-enabled point-of-care sensing using multiplexed paper-based sensors. *npj Digital Medicine* 2020; **3**: 1–8.

18      Holmström O, Linder N, Ngasala B, Mårtensson A, Linder E, Lundin M *et al.* Point-of-care mobile digital microscopy and deep learning for the detection of soil-transmitted helminths and Schistosoma haematobium. *Global Health Action* 2017; **10**: 1337325.

19      Joung H-A, Ballard ZS, Wu J, Tseng DK, Teshome H, Zhang L *et al.* Point-of-Care Serodiagnostic Test for Early-Stage Lyme Disease Using a Multiplexed Paper-Based Immunoassay and Machine Learning. *ACS Nano* 2020; **14**: 229–240.

20      Veli M, Ozcan A. Computational Sensing of Staphylococcus aureus on Contact Lenses Using 3D Imaging of Curved Surfaces and Machine Learning. *ACS Nano* 2018; **12**: 2554–2559.

21      Malkiel I, Mrejen M, Nagler A, Arieli U, Wolf L, Suchowski H. Plasmonic nanostructure design and characterization via Deep Learning. *Light: Science & Applications* 2018; **7**: 60.

22      Peurifoy J, Shen Y, Jing L, Yang Y, Cano-Renteria F, DeLacy BG *et al.* Nanophotonic particle simulation and inverse design using artificial neural networks. *Science Advances* 2018; **4**: eaar4206.

23      Liu D, Tan Y, Khoram E, Yu Z. Training Deep Neural Networks for the Inverse Design of Nanophotonic Structures. *ACS Photonics* 2018; **5**: 1365–1369.

24      Luo Y, Mengu D, Yardimci NT, Rivenson Y, Veli M, Jarrahi M *et al.* Design of task-specific optical systems using broadband diffractive neural networks. *Light: Science & Applications* 2019; **8**: 112.

25      Borhani N, Kakkava E, Moser C, Psaltis D. Learning to see through multimode fibers. *Optica, OPTICA* 2018; **5**: 960–966.

26      Shen Y, Harris NC, Skirlo S, Prabhu M, Baehr-Jones T, Hochberg M *et al.* Deep learning with coherent nanophotonic circuits. *Nature Photonics* 2017; **11**: 441–446.

27      Hughes TW, Williamson IAD, Minkov M, Fan S. Wave physics as an analog recurrent neural network. *Science Advances* 2019; **5**: eaay6946.

28    Miscuglio M, Mehrabian A, Hu Z, Azzam SI, George J, Kildishev AV et al. All-optical nonlinear activation function for photonic neural networks [Invited]. *Opt Mater Express, OME* 2018; **8**: 3851–3863.

29    Bueno J, Maktoobi S, Froehly L, Fischer I, Jacquot M, Larger L et al. Reinforcement learning in a large-scale photonic recurrent neural network. *Optica, OPTICA* 2018; **5**: 756–760.

30    Sande GV der, Brunner D, Soriano MC. Advances in photonic reservoir computing. *Nanophotonics* 2017; **6**: 561–576.

31    Hamerly R, Bernstein L, Sludds A, Soljačić M, Englund D. Large-Scale Optical Neural Networks Based on Photoelectric Multiplication. *Phys Rev X* 2019; **9**: 021032.

32    Lin X, Rivenson Y, Yardimci NT, Veli M, Luo Y, Jarrahi M et al. All-optical machine learning using diffractive deep neural networks. *Science* 2018; **361**: 1004.

33    Li J, Mengu D, Luo Y, Rivenson Y, Ozcan A. Class-specific differential detection in diffractive optical neural networks improves inference accuracy. *Adv Photonics* 2019; **1**: 046001.

34    Mengu D, Luo Y, Rivenson Y, Ozcan A. Analysis of Diffractive Optical Neural Networks and Their Integration With Electronic Neural Networks. *IEEE Journal of Selected Topics in Quantum Electronics* 2020; **26**: 1–14.

35    Marin-Palomo P, Kemal JN, Karpov M, Kordts A, Pfeifle J, Pfeiffer MHP et al. Microresonator-based solitons for massively parallel coherent optical communications. *Nature* 2017; **546**: 274–279.

36    Strickland D, Mourou G. Compression of amplified chirped optical pulses. *Optics Communications* 1985; **56**: 219–221.

37    Vabishchevich PP, Shcherbakov MR, Bessonov VO, Dolgova TV, Fedyanin AA. Femtosecond pulse shaping with plasmonic crystals. *Jetp Lett* 2015; **101**: 787–792.

38    Rahimi E, Şendur K. Femtosecond pulse shaping by ultrathin plasmonic metasurfaces. *J Opt Soc Am B, JOSAB* 2016; **33**: A1–A7.

39    Szipöcs R, Ferencz K, Spielmann C, Krausz F. Chirped multilayer coatings for broadband dispersion control in femtosecond lasers. *Opt Lett, OL* 1994; **19**: 201–203.

40    Supradeepa VR, Huang C-B, Leaird DE, Weiner AM. Femtosecond pulse shaping in two dimensions: Towards higher complexity optical waveforms. *Opt Express, OE* 2008; **16**: 11878–11887.

41    Weiner AM. Femtosecond pulse shaping using spatial light modulators. *Review of Scientific Instruments* 2000; **71**: 1929–1960.

42    Dugan MA, Tull JX, Warren WS. High-resolution acousto-optic shaping of unamplified and amplified femtosecond laser pulses. *J Opt Soc Am B, JOSAB* 1997; **14**: 2348–2358.

43    Weiner AM. Ultrafast optical pulse shaping: A tutorial review. *Optics Communications* 2011; **284**: 3669–3692.

44	Yelin D, Meshulach D, Silberberg Y. Adaptive femtosecond pulse compression. *Opt Lett, OL* 1997; **22**: 1793–1795.

45	Assion A, Baumert T, Bergt M, Brixner T, Kiefer B, Seyfried V *et al.* Control of Chemical Reactions by Feedback-Optimized Phase-Shaped Femtosecond Laser Pulses. *Science* 1998; **282**: 919–922.

46	Efimov A, Moores MD, Beach NM, Krause JL, Reitze DH. Adaptive control of pulse phase in a chirped-pulse amplifier. *Opt Lett, OL* 1998; **23**: 1915–1917.

47	Weiner AM, Leaird DE, Patel JS, Wullert JR. Programmable femtosecond pulse shaping by use of a multielement liquid-crystal phase modulator. *Opt Lett, OL* 1990; **15**: 326–328.

48	Bardeen CJ, Yakovlev VV, Wilson KR, Carpenter SD, Weber PM, Warren WS. Feedback quantum control of molecular electronic population transfer. *Chemical Physics Letters* 1997; **280**: 151–158.

49	Hillegas CW, Tull JX, Goswami D, Strickland D, Warren WS. Femtosecond laser pulse shaping by use of microsecond radio-frequency pulses. *Opt Lett, OL* 1994; **19**: 737–739.

50	Zeek E, Maginnis K, Backus S, Russek U, Murnane M, Mourou G *et al.* Pulse compression by use of deformable mirrors. *Opt Lett, OL* 1999; **24**: 493–495.

51	Divitt S, Zhu W, Zhang C, Lezec HJ, Agrawal A. Ultrafast optical pulse shaping using dielectric metasurfaces. *Science* 2019; **364**: 890–894.

52	Hashemi MR, Cakmakyapan S, Jarrahi M. Reconfigurable metamaterials for terahertz wave manipulation. *Rep Prog Phys* 2017; **80**: 094501.

53	Rahm M, Li J-S, Padilla WJ. THz Wave Modulators: A Brief Review on Different Modulation Techniques. *J Infrared Milli Terahz Waves* 2013; **34**: 1–27.

54	Danielson JR, Amer N, Lee Y-S. Generation of arbitrary terahertz wave forms in fanned-out periodically poled lithium niobate. *Appl Phys Lett* 2006; **89**: 211118.

55	Stepanov AG, Hebling J, Kuhl J. Generation, tuning, and shaping of narrow-band, picosecond THz pulses by two-beam excitation. *Opt Express, OE* 2004; **12**: 4650–4658.

56	Sato M, Higuchi T, Kanda N, Konishi K, Yoshioka K, Suzuki T *et al.* Terahertz polarization pulse shaping with arbitrary field control. *Nature Photonics* 2013; **7**: 724–731.

57	Keren-Zur S, Tal M, Fleischer S, Mittleman DM, Ellenbogen T. Generation of spatiotemporally tailored terahertz wavepackets by nonlinear metasurfaces. *Nature Communications* 2019; **10**: 1778.

58	Yongqian Liu, Sang-Gyu Park, Weiner AM. Terahertz waveform synthesis via optical pulse shaping. *IEEE Journal of Selected Topics in Quantum Electronics* 1996; **2**: 709–719.

59	Gingras L, Cooke DG. Direct temporal shaping of terahertz light pulses. *Optica, OPTICA* 2017; **4**: 1416–1420.

60      Yardimci NT, Jarrahi M. High Sensitivity Terahertz Detection through Large-Area Plasmonic Nano-Antenna Arrays. *Scientific Reports* 2017; **7**: 42667.

61      Cunningham PD, Valdes NN, Vallejo FA, Hayden LM, Polishak B, Zhou X-H *et al.* Broadband terahertz characterization of the refractive index and absorption of some important polymeric and organic electro-optic materials. *Journal of Applied Physics* 2011; **109**: 043505-043505–5.

62      Podzorov A, Gallot G. Low-loss polymers for terahertz applications. *Appl Opt, AO* 2008; **47**: 3254–3257.

63      Jin Y-S, Kim G-J, Jeon S-G. Terahertz dielectric properties of polymers. *Journal of the Korean Physical Society* 2006; **49**: 513–517.

64      Burghoff D, Kao T-Y, Han N, Chan CWI, Cai X, Yang Y *et al.* Terahertz laser frequency combs. *Nature Photonics* 2014; **8**: 462–467.

65      Bachmann D, Rösch M, Süess MJ, Beck M, Unterrainer K, Darmo J *et al.* Short pulse generation and mode control of broadband terahertz quantum cascade lasers. *Optica, OPTICA* 2016; **3**: 1087–1094.

66      Barbieri S, Ravaro M, Gellie P, Santarelli G, Manquest C, Sirtori C *et al.* Coherent sampling of active mode-locked terahertz quantum cascade lasers and frequency synthesis. *Nature Photonics* 2011; **5**: 306–313.

67      van der Weide DW. Delta-doped Schottky diode nonlinear transmission lines for 480-fs, 3.5-V transients. *Appl Phys Lett* 1994; **65**: 881–883.

68      Assefzadeh MM, Babakhani A. Broadband Oscillator-Free THz Pulse Generation and Radiation Based on Direct Digital-to-Impulse Architecture. *IEEE Journal of Solid-State Circuits* 2017; **52**: 2905–2919.

69      Abo-Bakr M, Feikes J, Holldack K, Kuske P, Peatman WB, Schade U *et al.* Brilliant, Coherent Far-Infrared (THz) Synchrotron Radiation. *Phys Rev Lett* 2003; **90**: 094801.

70      Partanen A, Väyrynen J, Hassinen S, Tuovinen H, Mutanen J, Itkonen T *et al.* Fabrication of terahertz wire-grid polarizers. *Appl Opt, AO* 2012; **51**: 8360–8365.

71      Kitahara H, Tsumura N, Kondo H, Takeda MW, Haus JW, Yuan Z *et al.* Terahertz wave dispersion in two-dimensional photonic crystals. *Phys Rev B* 2001; **64**: 045202.

72      Yardimci NT, Yang S-H, Berry CW, Jarrahi M. High-Power Terahertz Generation Using Large-Area Plasmonic Photoconductive Emitters. *IEEE Transactions on Terahertz Science and Technology* 2015; **5**: 223–229.

73      Kulce O, Mengu D, Rivenson Y, Ozcan A. All-Optical Information Processing Capacity of Diffractive Surfaces. *arXiv:200712813 [physics]* 2020.http://arxiv.org/abs/2007.12813 (accessed 9 Aug2020).

**Figures**

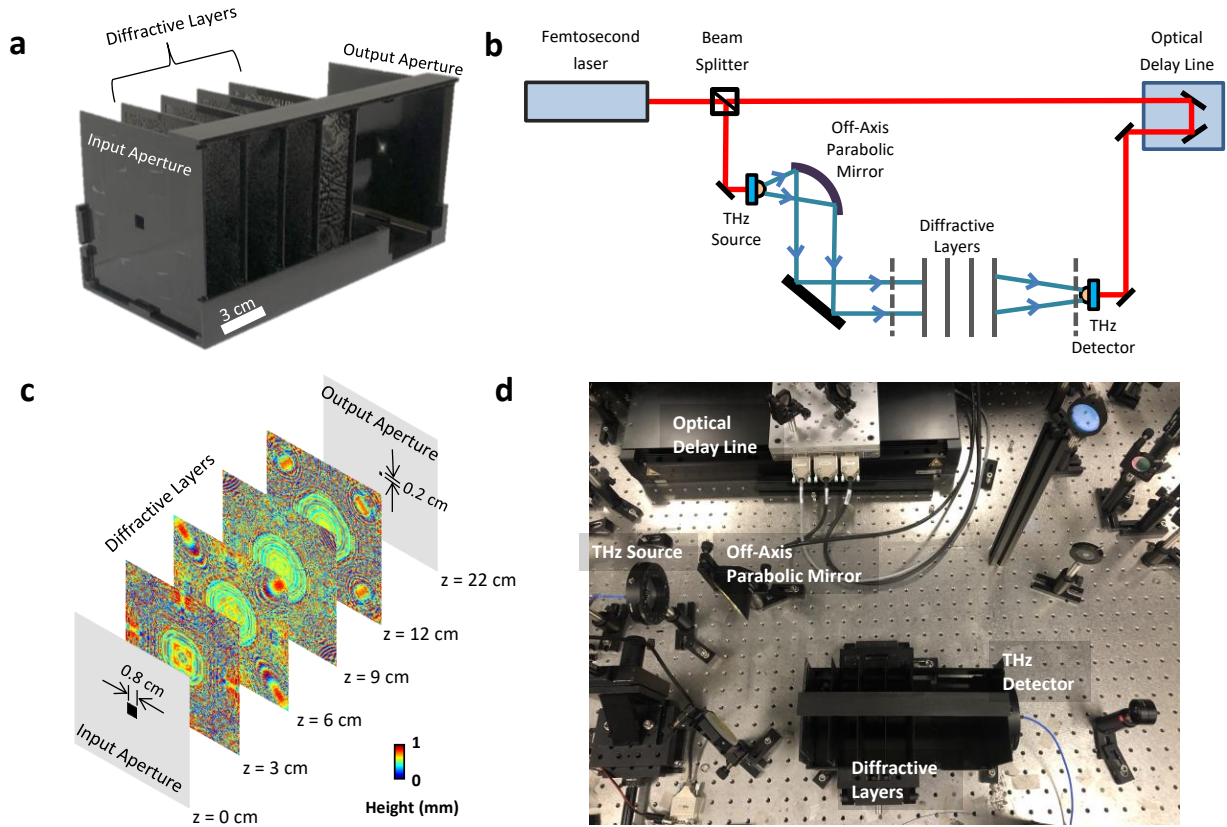

**Fig. 1 Schematic of the pulse shaping diffractive network and a photo of the experimental setup. a** 3D printed pulse shaping diffractive network that generates a square pulse with a width of 15.57 ps. **b** The schematic of the THz-TDS setup used in our experiments. The red line represents the optical path of a 780 nm femtosecond laser, and the blue line represents the terahertz beam. Dashed lines show the input and output apertures of the diffractive network. **c** The physical system layout of the pulse shaping diffractive network design. The input and output apertures are squares, with edge lengths of 0.8 cm and 0.2 cm, respectively. Gray regions on the aperture planes represent aluminum coating to block light transmission. **d** The photo of the experimental setup.

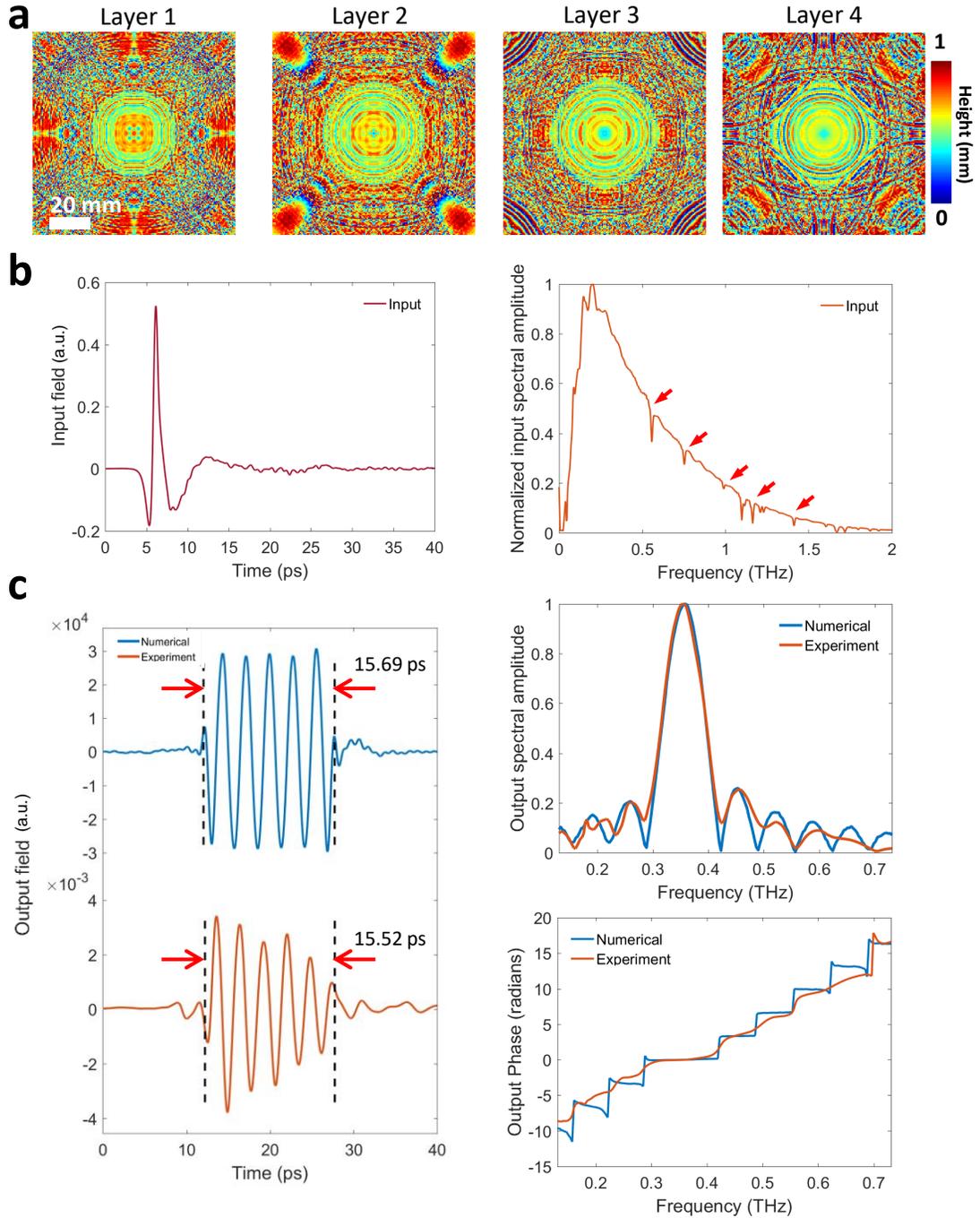

**Fig. 2 Pulse shaping diffractive network design and output results. a** The thickness profiles of the resulting diffractive layers after deep learning-based training in a computer. These diffractive layers synthesize a square pulse with a width of 15.69 ps over the output aperture for an input pulse shown in **b**. **b** Normalized input terahertz pulse measured right after the input aperture (see Fig. 1); in time-domain (left) and spectral domain (right). The red arrows on the measured spectral amplitude profile represent the water absorption bands at terahertz frequencies. **c Left**: The numerically computed (blue) and the experimentally measured (orange) output pulses in time domain. **Top right**: The normalized spectral amplitudes corresponding to the numerically computed (blue) and the experimentally measured (orange) pulses. **Bottom right**: Unwrapped spectral phase distributions computed based on the numerical forward model (blue) and the experimentally measured (orange) pulse.

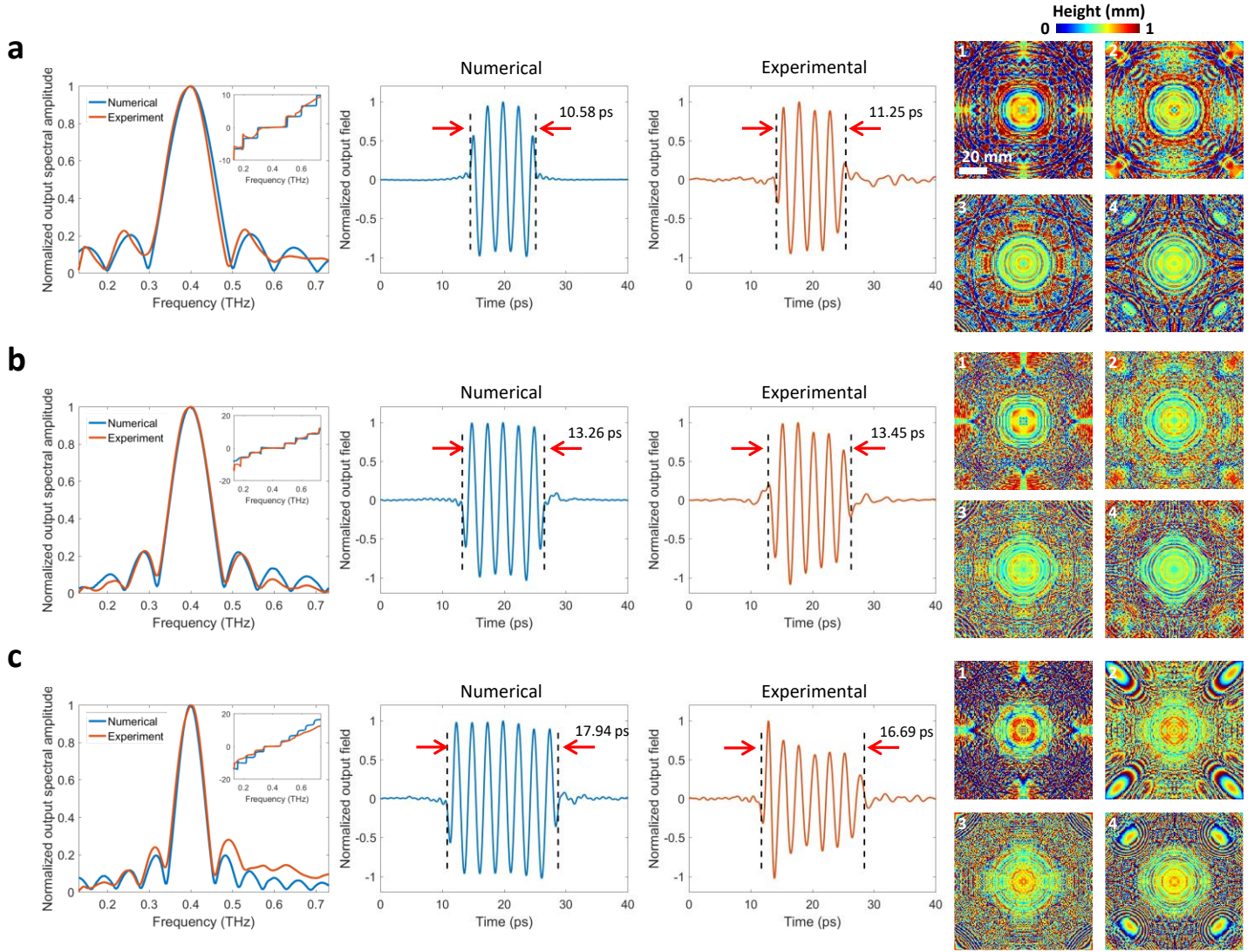

**Fig. 3 Experimental validation of different generic pulse shaping diffractive networks.** From left to right, the numerically computed (blue) and the experimentally measured (orange) normalized spectral amplitudes are illustrated with the inset plots showing the corresponding unwrapped spectral phase profiles; the numerically calculated (blue) normalized output pulse and the experimentally measured (orange) normalized output pulse are also shown along with the thickness profiles of the diffractive layers resulting from deep learning-based training for synthesizing the desired (ground truth) square pulses with pulse-widths of **(a)** 10.52 ps, **(b)** 13.02 ps and **(c)** 17.98 ps.

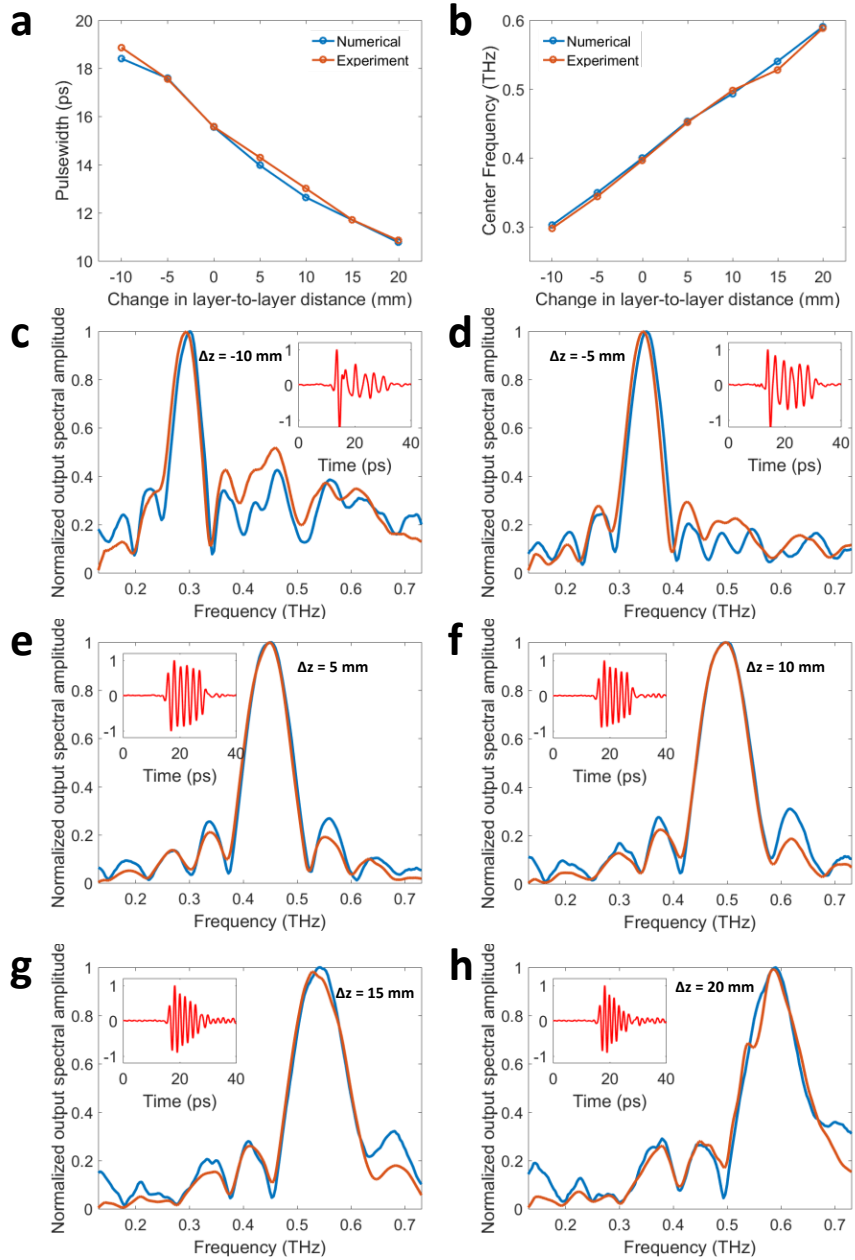

**Fig. 4 Pulse width tunability of diffractive networks. (a)** Numerically calculated and experimentally measured temporal pulse widths and **(b)** the corresponding shifts in the center frequency are depicted as a function of the inter-layer distances of a pulse shaping diffractive network that was originally trained for synthesizing a square pulse width of 15.50 ps (Δz = 0 mm, see Fig. 2). **(c-h)** The numerically computed (blue) and the experimentally measured (orange) normalized spectral amplitudes, with the inset plots showing the experimentally measured temporal waveform (red) when the layer-to-layer distances are changed by **(c)** Δz = -10 mm, **(d)** Δz = -5 mm, **(e)** Δz = 5 mm, **(f)** Δz = 10 mm, **(g)** Δz = 15 mm and **(h)** Δz = 20 mm. The negative (positive) sign indicates that the inter-layer axial distances decrease (increase).

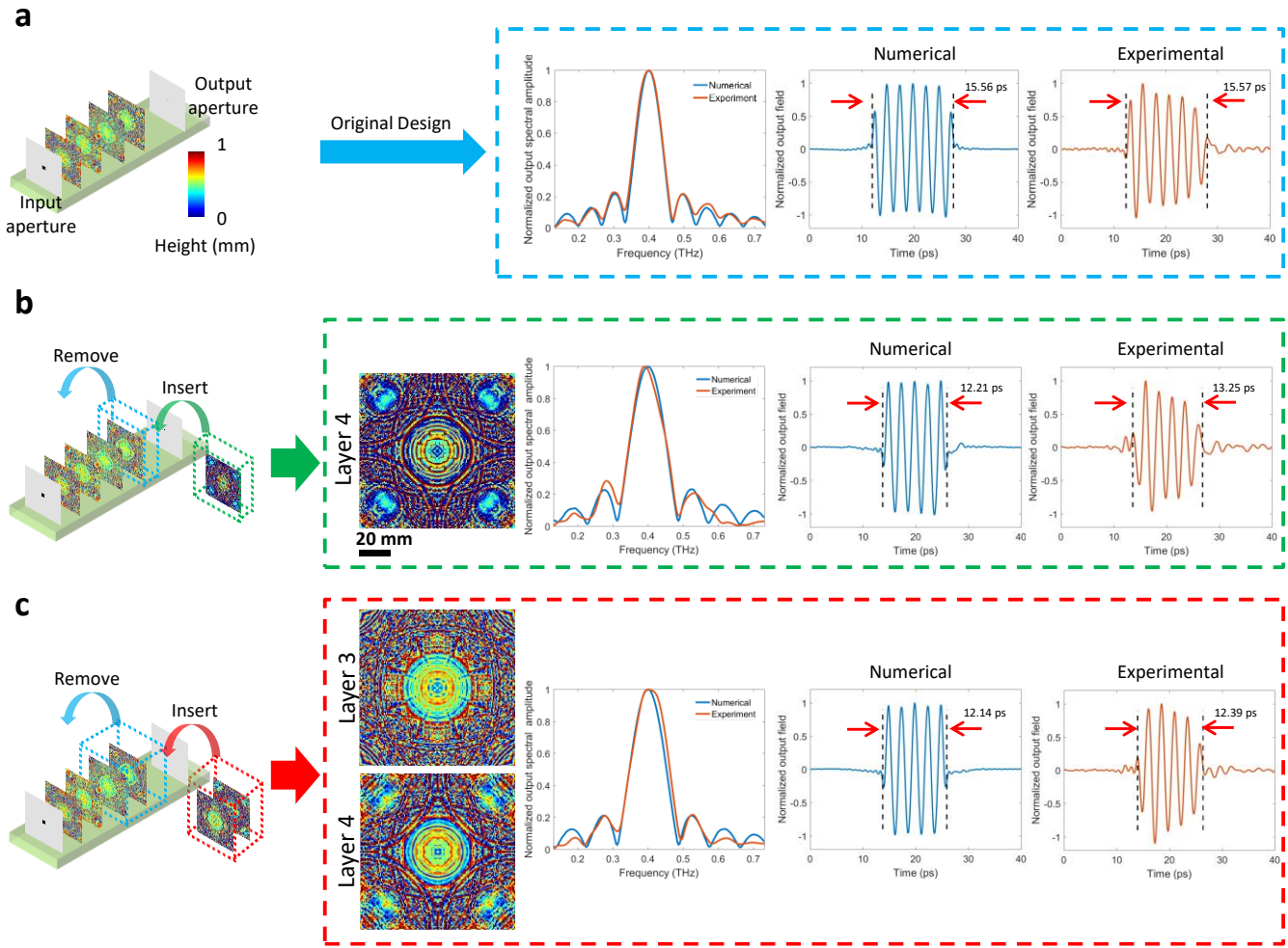

**Fig. 5 Changing the output temporal waveform of a diffractive network by a Lego-like transfer learning approach**. **a** The temporal and spectral output distributions (blue dashed box), synthesized by the original diffractive design that was trained to generate a 15.50 ps square pulse. **b** Replacing the last diffractive layer with another, newly trained diffractive layer to synthesize a 12.03 ps square pulse at the output. The thickness profile of the newly trained diffractive layer is shown, together with the normalized spectral and temporal profiles synthesized by this new diffractive network in green dashed box. **c** Replacing the last two diffractive layers with newly trained, two diffractive layers to synthesize a 12.03 ps square pulse at the output. The thickness profiles of the newly trained diffractive layers are shown, together with the normalized spectral and temporal profiles synthesized by this new diffractive network in red dashed box.

| | | 2 mm width output aperture | | 4 mm width output aperture | |
|---|---|---|---|---|---|
| | | **Efficiency** | | **Efficiency** | |
| Trained without absorption $L_p = \frac{\sum_\omega (I_{target} - I_{sensor})^2}{\sum_\omega I_{target}^2}$ | Test without absorption | **14.91%** | 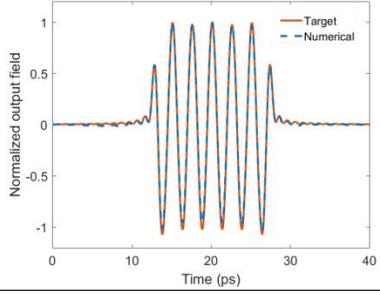 | **60.37%** | 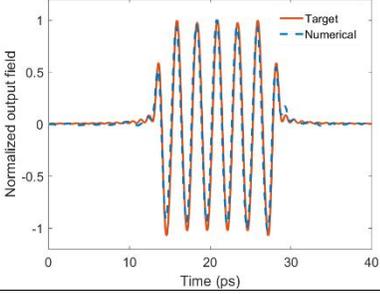 |
| | Test with absorption | **3.9%** | 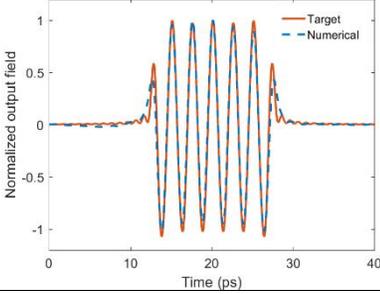 | **12%** | 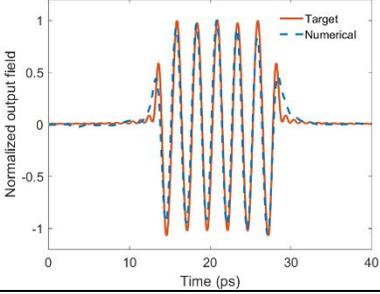 |
| Trained with absorption $L_p = \frac{\sum_\omega (I_{target} - I_{sensor})^2}{\sum_\omega I_{target}^2}$ | Test without absorption | **46.13%** | 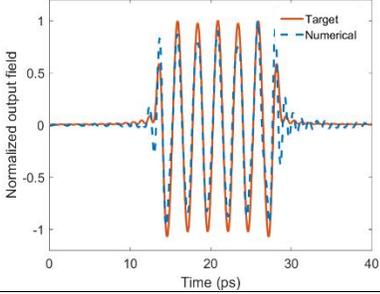 | **61%** | 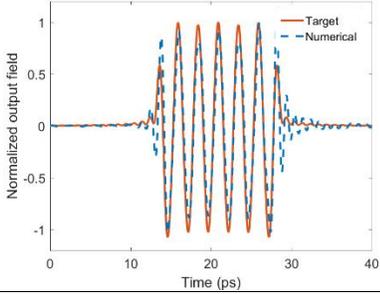 |
| | Test with absorption | **13.13%** | 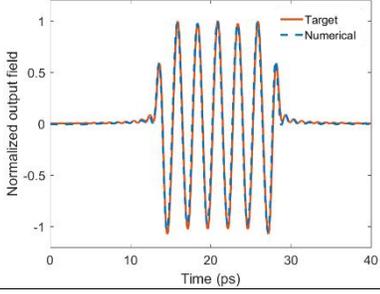 | **15.78%** | 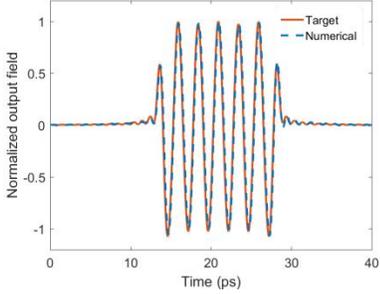 |
| Trained with absorption $L_p = \begin{cases} -\log\left(\frac{\eta}{\eta_{th}}\right), & \text{if } \eta < \eta_{th} \\ 0, & \text{if } \eta \geq \eta_{th} \end{cases}$ | Test without absorption | **31.14%** | 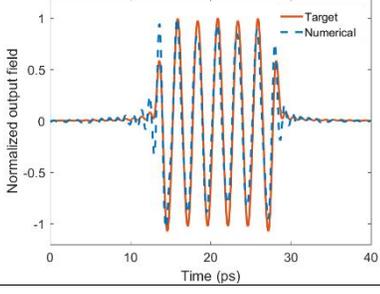 | **31.93%** | 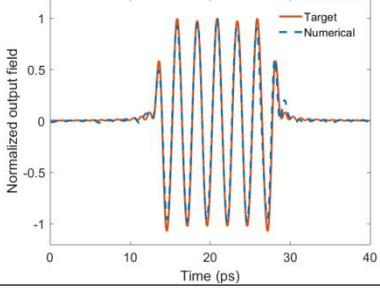 |
| | Test with absorption | **14.11%** | 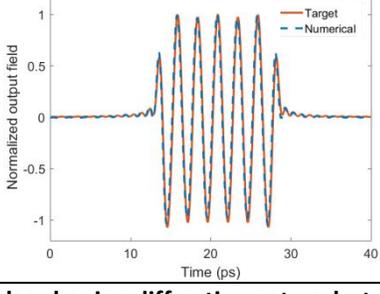 | **17.84%** | 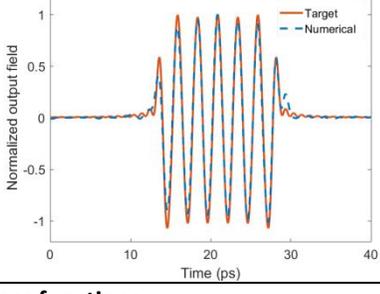 |

**Fig.6 Power efficiency values of pulse shaping diffractive networks trained with different loss functions.**